\documentclass{article}
\usepackage{amssymb}
\usepackage{amsfonts}
\usepackage{amsmath}
\usepackage[doublespacing]{setspace}

\newtheorem{theorem}{Theorem}

\newtheorem{proposition}[theorem]{Proposition}

\newenvironment{proof}[1][Proof]{\noindent\textbf{#1.} }{\ \rule{0.5em}{0.5em}}
\input{tcilatex}

\begin{document}

\title{Non-Hermitian $\mathcal{PT}$-symmetric and Hermitian Hamiltonians'
correspondence: Isospectrality and mass signature }
\author{Omar Mustafa$^{1}$ and S.Habib Mazharimousavi$^{2}$ \\
Department of Physics, Eastern Mediterranean University, \\
G Magusa, North Cyprus, Mersin 10,Turkey\\
$^{1}$E-mail: omar.mustafa@emu.edu.tr\\
\ $^{2}$E-mail: habib.mazhari@emu.edu.tr}
\maketitle

\begin{abstract}
A transformation of the form $x\rightarrow \pm iy\in i%
\mathbb{R}
;$ $x,y\in 
\mathbb{R}
$, or an equivalent similarity transformation with a metric operator $\eta $
are shown to map non-Hermitian $\mathcal{PT}$-symmetric Hamiltonians into
Hermitian partner Hamiltonians in Hilbert space. Isospectrality and mass
signature are also discussed.


\end{abstract}

\section{Introduction}

Recent developments on non-Hermitian Hamiltonians have documented that
Hermiticity is no more a necessary condition to secure the reality of the
spectrum [1-43]. Such developments are very much inspired by the nowadays
known as the Bender's and Boettcher's [1] conjecture in relaxing Hermiticity
condition and introducing the concept of $\mathcal{PT}$-symmetric quantum
mechanics (PTQM). Where, $\mathcal{P}$ denotes space reflection: $%
x\longrightarrow -x$ (i.e., parity operator) and $\mathcal{T}$ \ mimics the
time-reversal: $i\longrightarrow -i$. More specifically, \ if $\rho =%
\mathcal{PT}$ and $\rho \,H\,\rho ^{-1}=H$, then $H$ is $\mathcal{PT}$%
-symmetric. Moreover, if $\rho \,\Psi =\pm \Psi $ (i.e., $\Psi $ retains $%
\mathcal{PT}$-symmetry) the eigenvalues of a $\mathcal{PT}$-symmetric
Hamiltonian are real, otherwise the eigenvalues come out in
complex-conjugate pairs (a phenomenon known as spontaneous breakdown of $%
\mathcal{PT}$-symmetry).

Such a PTQM theory, nevertheless, has stimulated intensive research on the
non-Hermitian Hamiltonians and led to the so-called pseudo-Hermitian
Hamiltonians (i.e., Hamiltonians satisfying $\xi \,H\,\xi ^{-1}=H^{\dagger }$
or $\xi \,H=H^{\dagger }\,\xi $, where $\xi $ is a Hermitian invertible
linear operator and $(^{\dagger })$ denotes the adjoint) by Mostafazadeh
[20-25] which form a broader class of non-Hermitian Hamiltonians with real
spectra that encloses within those $\mathcal{PT}$-symmetric ones. Moreover,
not restricting $\xi $ to be Hermitian (cf., e.g., Bagchi and Quesne [38]),
and linear and/or invertible (cf., e.g., Solombrino [32], Fityo [33], and
Mustafa and Mazharimousavi [34-37]) would lead to real spectra.

In the process, on the other hand, some quantum mechanical models of certain
exceptional $\mathcal{PT}$-symmetric complex interactions, i.e., a $\mathcal{%
PT}$-symmetric potential satisfies 
\begin{equation}
\mathcal{PT}V\left( x\right) =V\left( x\right) \iff V\left( x\right) =\left[
V\left( -x\right) \right] ^{\ast },
\end{equation}%
just happen to have their partners that are strictly equivalent to real
potentials after being exposed to some supersymmetric quantum mechanical
treatment [11] or integral, Fourier-like transformation [12]. Jones and
Mateo [4] have, moreover, used a Darboux-type similarity transformation and
have shown that for the Bender's and Boettcher's [1] non-Hermitian $\mathcal{%
PT}$-symmetric Hamiltonian $H=p^{2}-g\left( ix\right) ^{N};$ $N=4$, there
exists an equivalent Hermitian Hamiltonian $h=\sigma ^{-1}H\sigma $; $\sigma
=\exp \left( Q/2\right) $, where $\sigma $ is Hermitian and positive
definite. Similar proposal was carried out by Bender et al. [3]. For more
details the reader is advised to refer to [3,4]. In our current methodical
proposal, we try to have our input in this direction and fill this gap
partially, at least.

Through the forthcoming proposition (in section 2) or through a similarity
transformation (in section 3) with a metric operator $\eta $ (defined in
(21) below) we report that for every non-Hermitian complex $\mathcal{PT}$%
-symmetric Hamiltonian (with positive mass $m=m_{+}=+\left\vert m\right\vert 
$) there exists a Hermitian partner Hamiltonian (with negative mass $%
m=m_{-}=-\left\vert m\right\vert $) in Hilbert space $L^{2}\left( 
\mathbb{R}
\right) =\mathcal{H}$. In section 3, we also discuss isospectrality and
orthonormalization conditions associated with both the Hermitian partner
(not necessarily $\mathcal{PT}$-symmetric) and the non-Hermitian $\mathcal{PT%
}$-symmetric Hamiltonians. An obvious correspondence is constructed,
therein. This has not been discussed elsewhere, to the best of our
knowledge. We give our concluding remarks in section 4.

\section{A transformation toy: $x\longrightarrow \pm iy\,;\,\,x,\,y\in 
\mathbb{R}
$}

In connection with an over simplified transformation toy $x\longrightarrow
\pm iy\,\in i%
\mathbb{R}
;\,\,x,\,y\in 
\mathbb{R}
$ ($x\longrightarrow \pm iy$ to be understood as $x\longrightarrow +iy$
and/or $x\longrightarrow -iy$), t' Hooft and Nobbenhuis [44] have used a
complex space-time symmetry transformation

\begin{equation*}
x\longrightarrow iy\Longleftrightarrow \,p_{x}\rightarrow
-ip_{y}\,;\,\,x,\,y\in 
\mathbb{R}
,
\end{equation*}%
between de-Sitter and anti-de-Sitter space to identify vacuum solutions with
zero cosmological constant (used later on by Assis and Fring [45] to provide
a simple proof of the reality of the spectrum of $p^{2}+z^{2}\left(
iz\right) ^{2m+1}$). However, in their instructive harmonic oscillator [44]
example%
\begin{equation}
H_{x}=\frac{p_{x}^{2}}{2m}+\frac{1}{2}m\omega ^{2}x^{2}=\omega \left(
a_{x}^{\dagger }a_{x}+\frac{1}{2}\right) ,
\end{equation}%
with the annihilation and creation operators%
\begin{equation}
a_{x}=\sqrt{\frac{1}{2m\omega }}\left( m\omega x+ip_{x}\right) \,,\text{ \ \
\ }a_{x}^{\dagger }=\sqrt{\frac{1}{2m\omega }}\left( m\omega x-ip_{x}\right)
,
\end{equation}%
they have shown (using $x\longrightarrow iy,\,p_{x}\rightarrow -ip_{y}$)
that the corresponding Hamiltonian reads%
\begin{equation}
H_{y}=\omega \left( a_{y}^{\dagger }a_{y}-\frac{1}{2}\right) ,
\end{equation}%
with%
\begin{equation}
a_{y}=\sqrt{\frac{1}{2m\omega }}\left( m\omega y+ip_{y}\right) ,\text{ \ }%
a_{y}^{\dagger }=\sqrt{\frac{1}{2m\omega }}\left( -m\omega y+ip_{y}\right) .
\end{equation}%
Under such settings, $H_{x}\longrightarrow -H_{y}$ and whilst the
eigenvalues of $H_{x}$ are $\left[ \omega \left( n+1/2\right) \right] $
those of $H_{y}$ read $\left[ -\omega \left( n+1/2\right) \right] $.
Consequently, the ground state $\sim \exp \left( -m\omega x^{2}/2\right) $
in the $x$-space is normalizable, whereas the ground state $\sim \exp \left(
+m\omega y^{2}/2\right) $ in the $y$-space is non-normalizable.

Within similar spiritual lines, Tanaka [46] has shown that a transformation
of the form%
\begin{equation}
x\in 
\mathbb{R}
\longrightarrow -iy\in i%
\mathbb{R}
\,,\text{ \ \ }p_{x}\longrightarrow ip_{y}\in i%
\mathbb{R}%
\end{equation}%
would map a non-Hermitian $PT$-symmetric potential $V\left( x\right) \in 
\mathbb{C}
$\ (or any non-Hermitian $PT$-symmetric function $f\left( x\right) \in 
\mathbb{C}
$ in general, so to speak) into a Hermitian (but not necessarily $PT$%
-symmetric) potential $V\left( y\right) \in 
\mathbb{R}
$. The proof of which is straightforward. Using equation (1), one would
write (with $z=-iy$ for simplicity of notations)%
\begin{equation}
V\left( x\right) \mid _{x\rightarrow z}=\left[ V\left( -x\right) \right]
^{\ast }\mid _{x\rightarrow z}=V^{\ast }\left( -z^{\ast }\right) .
\end{equation}%
This would in turn imply that%
\begin{equation}
V\left( -iy\right) =V^{\ast }\left( -iy\right) \in 
\mathbb{R}
\Longrightarrow V\left( y\right) =V^{\ast }\left( y\right) \in 
\mathbb{R}
,
\end{equation}%
where $V\left( y\right) $ is a real-valued function, therefore.\ Some
illustrative examples can be found section 6 of [46].

In this respect, a remedy for the t' Hooft and Nobbenhuis [44] harmonic
oscillator Hamiltonian above may be sought in a mass parametrization recipe
accompanied with the de-Sitter and anti-de-Sitter transformation $x\in 
\mathbb{R}
\longrightarrow iy\in i%
\mathbb{R}
$. That is, 
\begin{equation}
m=m_{\pm }\Longrightarrow m=\left\{ 
\begin{tabular}{l}
$m_{+}=+\left\vert m\right\vert >0$ \\ 
$m_{-}=-\left\vert m\right\vert <0$%
\end{tabular}%
\right.
\end{equation}%
Such a mass parametrization would, in turn, suggest that the t' Hooft and
Nobbenhuis [44] harmonic oscillator%
\begin{equation}
H_{x}=H_{x;m_{+}}=\frac{p_{x}^{2}}{2m_{+}}+\frac{1}{2}m_{+}\omega ^{2}x^{2}
\end{equation}%
in (2) with $m_{+}=-m_{-}$ reads%
\begin{equation}
H_{y;m_{-}}=\frac{p_{y}^{2}}{2m_{-}}+\frac{1}{2}m_{-}\omega ^{2}y^{2}\text{ }%
\in L^{2}\left( 
\mathbb{R}
\right) =\mathcal{H}\text{.}
\end{equation}%
In this case both $H_{x;m_{+}}$ and $H_{y;m_{-}}$ are isospectral and both
admit normalizable eigenfunctions. For example, the ground state in $x$%
-space $\sim \exp \left( -m_{+}\omega x^{2}/2\right) $ and that in the $y$%
-space $\sim \exp \left( -m_{-}\omega y^{2}/2\right) $ are both
normalizable. The mass parametrization recipe does the trick, therefore.

Such observations would unavoidably manifest the following proposition.

\begin{proposition}
\emph{For every non-Hermitian complex }$\mathcal{PT}$\emph{-symmetric
Hamiltonian with positive mass (i.e., }$m=m_{+}=+\left\vert m\right\vert $)%
\emph{\ there exists a Hermitian (but not necessarily isospectral neither
necessarily }$\mathcal{PT}$-symmetric\emph{) partner Hamiltonian with
negative mass (i.e., }$m=m_{-}=-\left\vert m\right\vert $\emph{) in Hilbert
space }$L^{2}\left( 
\mathbb{R}
\right) =\mathcal{H}$\emph{.}
\end{proposition}

\begin{proof}
Let%
\begin{equation}
H_{x;m_{+}}=\frac{p_{x}^{2}}{2m_{+}}+V\left( x;m_{+}\right) \text{ };\text{
\ }V\left( x;m_{+}\right) =V^{\ast }\left( -x;m_{+}\right) \in 
\mathbb{C}
,
\end{equation}%
be a non-Hermitian complex $\mathcal{PT}$-symmetric Hamiltonian (with $%
m=+\left\vert m\right\vert $) with a corresponding $\mathcal{PT}$-symmetric
eigenfunctions $\Psi \left( x;m_{+}\right) $ such that 
\begin{equation*}
H_{x;m_{+}}\Psi \left( x;m_{+}\right) =E_{m_{+}}\Psi \left( x;m_{+}\right) .
\end{equation*}%
Then a mapping of the sort%
\begin{equation}
x\in 
\mathbb{R}
\longrightarrow \pm iy\in i%
\mathbb{R}
\Longleftrightarrow \text{ \ }p_{x}\longrightarrow \mp ip_{y}\in i%
\mathbb{R}
\,;\,\,x,\,y\in 
\mathbb{R}
,
\end{equation}%
would imply%
\begin{equation}
H_{x;m_{+}}\Psi \left( x;m_{+}\right) =E_{m_{+}}\Psi \left( x;m_{+}\right)
\Longleftrightarrow H_{y;m_{-}}\Phi \left( y;m_{-}\right) =E_{m_{-}}\Phi
\left( y;m_{-}\right) ,
\end{equation}%
where the substitution $m_{+}=-m_{-}$ is used and%
\begin{equation}
H_{y;m_{-}}=\frac{p_{y}^{2}}{2m_{-}}+V\left( y;m_{-}\right) \text{ }\in
L^{2}\left( 
\mathbb{R}
\right) ;\text{ \ }V\left( y;m_{-}\right) =V^{\ast }\left( y;m_{-}\right)
\in 
\mathbb{R}
.
\end{equation}%
which is Hermitian (\emph{but not necessarily isospectral with }$H_{x;m_{+}}$%
\emph{\ of (12) neither necessarily }$\mathcal{PT}$-symmetric). QED.
\end{proof}

Illustrative examples are ample. In the complex "shifted by an imaginary
constant" $\mathcal{PT}$-symmetric oscillator Hamiltonian (cf., e.g.,
Mustafa and Znojil [18]) a companied by a properly regularized
attractive/repulsive core (with the mass term kept intact) 
\begin{equation}
H_{x;m_{+}}=\frac{p_{x}^{2}}{2m_{+}}+V\left( x;m_{+}\right) =\frac{p_{x}^{2}%
}{2m_{+}}+\frac{m_{+}\omega ^{2}}{2}\left( x-ic\right) ^{2}+\frac{G\left(
m_{+},\alpha \right) }{\left( x-ic\right) ^{2}},
\end{equation}%
would, under the transformation $x\longrightarrow \pm iy$ and with%
\begin{equation*}
G\left( m_{+},\alpha \right) =\frac{\hslash ^{2}\left( \alpha
^{2}-1/4\right) }{2m_{+}},
\end{equation*}%
imply%
\begin{equation}
V\left( y;m_{+}\right) =-\frac{m_{+}\omega ^{2}}{2}\left( y\mp c\right) ^{2}-%
\frac{\hslash ^{2}}{2m_{+}}\frac{\left( \alpha ^{2}-1/4\right) }{\left( y\mp
c\right) ^{2}}\in 
\mathbb{R}
.
\end{equation}%
Which is not only real valued but also $\mathcal{PT}$-symmetric (with parity
performing reflection about $y=\pm c$ rather than $y=0$). In such a case,%
\begin{equation*}
H_{x;m_{+}}\longrightarrow H_{y;m_{-}}=-H_{y;m_{+}}\in L^{2}\left( 
\mathbb{R}
\right) =\mathcal{H}
\end{equation*}%
where%
\begin{eqnarray}
H_{y;m_{-}} &=&\frac{P_{y}^{2}}{2m_{-}}+V\left( y;m_{-}\right)  \notag \\
&=&\frac{P_{y}^{2}}{2m_{-}}+\frac{m_{-}\omega ^{2}}{2}\left( y\mp c\right)
^{2}+\frac{\hslash ^{2}}{2m_{-}}\frac{\left( \alpha ^{2}-1/4\right) }{\left(
y\mp c\right) ^{2}}.
\end{eqnarray}%
Obviously, $H_{y;m_{-}}$ is not only Hermitian but also $\mathcal{PT}$%
-symmetric and shares the same eigenvalues with $H_{x;m_{+}}$ in (16), i.e.,%
\begin{equation*}
E_{n}=E_{m_{+}}=E_{m_{-}}=\left\{ 
\begin{tabular}{l}
$2n+1;$ for $\alpha =\pm 1/2,$ \\ 
$4n+2+2q\alpha ;$ otherwise,%
\end{tabular}%
\right. \,n=0,1,\cdots ,
\end{equation*}%
where $q=\pm 1$ denotes quasi-parity. Obviously the spectrum remains
discrete, real, and bounded from below and the wave functions remain
normalizable (cf., e.g., Znoijl [7] for more details), with a $c$ shift of
the coordinate up or down.

Moreover, the Bender's and Boettcher's [1] non-Hermitian $\mathcal{PT}$%
-symmetric Hamiltonian, with the $\mathcal{PT}$-symmetric potential $V\left(
x\right) =-g\left( ix\right) ^{\nu }\in 
\mathbb{C}
;\nu ,g\in 
\mathbb{R}
,\ \nu \geq 2,$ $g>0$,%
\begin{equation}
H_{x;m_{+}}=\frac{p_{x}^{2}}{2m_{+}}+V\left( x;m_{+}\right) =\frac{p_{x}^{2}%
}{2m_{+}}-g\left( ix\right) ^{\nu },
\end{equation}%
would, under the transformation $x\longrightarrow -iy$, yield%
\begin{equation}
H_{y;m_{-}}=\frac{P_{y}^{2}}{2m_{-}}+V\left( y;m_{-}\right) =\frac{P_{y}^{2}%
}{2m_{-}}-g\left( y\right) ^{\nu },
\end{equation}%
which is Hermitian (but non-$\mathcal{PT}$-symmetric for odd $\nu $ and $%
\mathcal{PT}$-symmetric for even $\nu $, i.e., conditional $\mathcal{PT}$%
-symmetric). Whilst Hermiticity is secured in the partner Hermitian
Hamiltonian, the boundary conditions and normalizability are not. Therefore
isospectrality is a different issue that remains "to-be-determined" and to
be partially discussed below.

The above were just few of the many examples available in the literature
where their non-Hermitian $\mathcal{PT}$-symmetric Hamiltonians find
Hermitian partners in the regular Hilbert space. Whenever one encounters
such cases, the possibility of isospectrality should always be tested in the
process. In the light of the above proposition, we may observe that our
simple transformation toy $x\in 
\mathbb{R}
\longrightarrow \pm iy\in i%
\mathbb{R}
$, could be interpreted as a counterclockwise/clockwise rotation by $\theta
=\pm \pi /2$ of the full real $x$-axis and would, effectively, just map a
point $z_{1}=x$ into a point $z_{2}=\pm iy$ \ on the imaginary $y$-axis of
the complex $z$- plane.

\section{A similarity transformation toy: isospectrality and mass signature}

In the search for a more technical metric operators' language, one may very
well use Ben-Aryeh's and Barak's [5] similarity transformation with a metric
operator 
\begin{equation}
\eta =\exp \left( -i\beta \,x\partial _{x}\right) \,;\text{ \ }\beta ,x\in 
\mathbb{R}
.
\end{equation}%
Which transforms a power series%
\begin{equation}
F\left( x\right) =\sum\limits_{n=0}^{\infty }A_{n}\,x^{n}\in 
\mathbb{R}
,
\end{equation}%
into%
\begin{equation}
G\left( x\right) =\eta \,F\left( x\right) \,\eta
^{-1}=\sum\limits_{n=0}^{\infty }A_{n}\,\left( e^{-i\beta }x\right) ^{n}\in 
\mathbb{C}
.
\end{equation}%
Where $G\left( x\right) $ is a non-Hermitian $\mathcal{PT}$-symmetric
function and satisfies the similarity transformation relation $\eta
\,^{-1}G\left( x\right) \,\eta =F\left( x\right) \in 
\mathbb{R}
$. To reflect such a result onto the transformation toy in the above section
we choose \ $\beta =\pm \pi /2$. This immediately mandates that a
non-Hermitian $\mathcal{PT}$-symmetric Hamiltonian $H_{\mathcal{PT}}$ can be
mapped into its partner Hermitian Hamiltonian $H$ (\emph{but not necessarily
isospectral neither necessarily }$\mathcal{PT}$-symmetric) through a
similarity transformation%
\begin{equation}
\eta ^{-1}H_{\mathcal{PT}}\,\eta =H\iff H_{\mathcal{PT}}=\eta H\eta ^{-1}.
\end{equation}%
Where%
\begin{equation}
H_{\mathcal{PT}}=\frac{p_{x}^{2}}{2m_{+}}+V\left( x;m_{+}\right) \,;\text{ }%
V\left( x;m_{+}\right) =\left[ V\left( -x;m_{+}\right) \right] ^{\ast }\in 
\mathbb{C}
,
\end{equation}%
and%
\begin{equation}
H=\frac{p_{x}^{2}}{2m_{-}}+V\left( ix;m_{-}\right) \text{ };\text{ \ }%
V\left( ix;m_{-}\right) =V\left( x;m_{-}\right) =V^{\ast }\left(
x;m_{-}\right) \in 
\mathbb{R}
.
\end{equation}%
$H$ denotes the Hermitian partner Hamiltonian in Hilbert space with real
eigenvalues, therefore.

Under such settings, one can easily show that $\eta x\eta ^{-1}=\pm ix$
(i.e., $x\rightarrow \pm ix$, which practically imitates our original
transformation toy above) and consequently a non-Hermitian $\mathcal{PT}$%
-symmetric potential $V_{\mathcal{PT}}\left( x\right) $ would be transformed
into its real-valued (by the virtue of equation (2)) partner potential $%
V\left( \pm ix;m_{+}\right) \in 
\mathbb{R}
$ through the relation 
\begin{equation}
\eta ^{-1}V_{\mathcal{PT}}\left( x\right) \eta =\eta ^{-1}V\left(
x;m_{+}\right) \eta =V\left( \pm ix;m_{+}\right) =\left[ V\left( \pm
ix;m_{+}\right) \right] ^{\ast }\in 
\mathbb{R}
.
\end{equation}

On the other hand, the proof of the related isospectrality between $H_{%
\mathcal{PT},m_{+}}$ in (25) and its Hermitian partner Hamiltonian $%
H_{m_{-}} $ in (26) seems to be a straightforward one. Let $E_{n,m_{+}}$ and 
$\Psi _{n}\left( x;m_{+}\right) $ be the eigenvalues and eigenfunctions of
the complex $\mathcal{PT}$-symmetric Hamiltonian $H_{\mathcal{PT},m_{+}}$,
respectively, then 
\begin{eqnarray*}
H_{\mathcal{PT},m_{+}}\Psi _{n}\left( x;m_{+}\right) &=&E_{n,m_{+}}\Psi
_{n}\left( x;m_{+}\right) \implies \\
\eta ^{-1}\eta H_{m_{+}}\left[ \eta ^{-1}\Psi _{n}\left( x;m_{+}\right) %
\right] &=&E_{n,m_{+}}\left[ \eta ^{-1}\Psi _{n}\left( x;m_{+}\right) \right]
\implies \\
H_{m_{+}}\Phi _{n}\left( x;m_{+}\right) &=&E_{n,m_{+}}\Phi _{n}\left(
x;m_{+}\right) \Longrightarrow
\end{eqnarray*}%
\begin{equation}
H_{m_{-}}\Phi _{n}\left( x;m_{-}\right) =E_{n,m_{-}}\Phi _{n}\left(
x;m_{-}\right) ,
\end{equation}%
where $\eta ^{-1}\Psi _{n}\left( x;m_{+}\right) =\Phi _{n}\left(
x;m_{+}\right) \in L^{2}\left( 
\mathbb{R}
\right) $ are the eigenfunctions for $H_{m_{+}}$ in the Hilbert space. Both
the non-Hermitian complex $\mathcal{PT}$-symmetric Hamiltonian $H_{\mathcal{%
PT},m_{+}}$ \ and its Hermitian partner Hamiltonian $H_{m_{+}}$ are
isospectral, therefore. Under such settings, we may observe that our
examples in the previous section fit into such isospecrtrality argument, no
doubt.

However, an immediate example on "temporary-fragile-isospectrality" may be
sought in the complex $\mathcal{PT}$-symmetric potential $V\left( x\right)
=-A\func{sech}^{2}\left( x-ic\right) $. Which upon the de-Sitter
anti-de-Sitter transformation would be mapped into $V\left( x\right)
=-A/\cos ^{2}x$ that manifests an unbounded spectrum because of the negative
sign. Nonetheless, an immediate remedy may be sought in the parametrization
of the coupling parameter, i.e., $A\longrightarrow -B\in 
\mathbb{R}
$ (in analogy with the mass parametrization in (9)). This would, in turn,
take $V\left( x\right) =-A/\cos ^{2}x$ (which does not support bound states)
into $V\left( x\right) =B/\cos ^{2}x\in 
\mathbb{R}
$ (which supports bound states).

In due course, we find that the complex $\mathcal{PT}$-symmetric
Hamiltonians find their Hermitian partners in the regular Hilbert space
through either a simple transformation toy (13) or a similarity
transformation toy (23) \ accompanied by a mass parametrization recipe (9)
and/or an analogous coupling constant recipe is an unavoidable conclusion.

Nevertheless, having had established this fact, we may now try to explore
the orthonormalization conditions. Since $\Phi _{n}\left( x\right) \in
L^{2}\left( 
\mathbb{R}
\right) $ are the eigenfunctions for $H$ in Hilbert space, they satisfy the
regular quantum mechanical orthonormalization condition%
\begin{equation}
\left\langle \Phi _{k}\left( x\right) \left\vert \Phi _{n}\left( x\right)
\right. \right\rangle =\delta _{kn}.
\end{equation}%
Consequently, the established connection $\Phi _{n}\left( x\right) =\eta
^{-1}\Psi _{n}\left( x\right) \in L^{2}\left( 
\mathbb{R}
\right) $ would imply%
\begin{equation}
\left\langle \eta ^{-1}\Psi _{k}\left( x\right) \left\vert \eta ^{-1}\Psi
_{n}\left( x\right) \right. \right\rangle =\delta _{kn}.
\end{equation}%
Which in turn yields%
\begin{equation}
\left\langle \Psi _{k}\left( x\right) \left\vert \mp i\eta ^{-1}\left[ \eta
^{-1}\Psi _{n}\left( x\right) \right] \right. \right\rangle =\delta
_{kn}\iff \left\langle \Psi _{k}\left( x\right) \left\vert \Psi _{n}\left(
-x\right) \right. \right\rangle =\pm i\delta _{kn}.
\end{equation}%
An obvious and immediate correspondence between the regular quantum
mechanical orthonormalization condition (20) and that associated with the
non-Hermitian complex $\mathcal{PT}$-symmetric Hamiltonians (22) is
constructed, therefore. However, we could not find any example that may
satisfy such a condition. The orthonormalizable set of wave functions
satisfying this condition is an empty set. This should be anticipated since
the normalizable wave functions of the Hermitian Hamiltonians are not
expected to be safely transformed (along with the associated well-defined
boundary conditions in Hilbert space) into the complex space.

\section{Concluding remarks}

In this work, we have introduced a simple transformation, $x\longrightarrow
\pm iy\,;\,\,x,\,y\in 
\mathbb{R}
$, that allowed non-Hermitian $\mathcal{PT}$-symmetric Hamiltonians to find
their Hermitian (\emph{but not necessarily isospectral neither necessarily }$%
\mathcal{PT}$-symmetric) partners in Hilbert space. We have also introduced
a similarity transformation recipe (with a metric operator $\eta $ in (21))
that proved to provide a more mathematical accessibility to the
orthonormalization conditions associated with both the Hermitian (not
necessarily $\mathcal{PT}$-symmetric) and the non-Hermitian $\mathcal{PT}$%
-symmetric (not necessarily isospectral) Hamiltonians.

Moreover, the parametrized-mass signature (an almost forgotten and usually
deliberately dismissed for the sake of mathematical manipulation simplicity)
is shown to play a significant role in the current methodical proposal. An
analogous coupling parameter's recipe is shown to play a similar role as
that of the parametrized mass. Yet, within the lines of the later, Znojil
[43] in his mass-sign duality proposal, has observed that the non-Hermitian
cubic oscillator's Hamiltonians $H_{\pm }=p^{2}\pm m^{2}x^{2}+ifx^{3}$ with
opposite sign mass signatures are (up to a constant shift) isospectral. For
the feasibly significant role it may play, the mass \ term should always be
kept intact with the associated Hamiltonians, therefore.

Finally, as long as our non-Hermitian $\mathcal{PT}$-symmetric Hamiltonians $%
H_{\mathcal{PT}}$ find their Hermitian partners (\emph{not necessarily
isospectral neither necessarily }$\mathcal{PT}$-symmetric) in the Hilbert
space (where boundary conditions and consequently orthonormalizability are
feasibly very well defined), either through a simple transformation toy (13)
or a similarity transformation toy (23) \ accompanied by a mass
parametrization recipe (9) and/or an analogous coupling constant recipes,
the non-Hermitian $\mathcal{PT}$-symmetric quantum mechanics remains safe
and deserves to be advocated irrespective with the orthodoxal mathematical
(though rather fragile) Hermiticity requirement.\emph{\newpage }

\end{document}